\documentclass[aps,prd,showpacs,twocolumn,nofootnoteinbib,groupedaddress]{revtex4}
\usepackage{epsfig,amssymb,amsmath,latexsym}
\newcommand\al{\alpha}
\newcommand\be{\beta}

\newcommand\beq{\begin{equation}}
\newcommand\eeq{\end{equation}}
\newcommand\bea{\begin{eqnarray}}
\newcommand\eea{\end{eqnarray}}
\newcommand\bi{\begin{itemize}}
\newcommand\ei{\end{itemize}}
\newcommand\ben{\begin{enumerate}}
\newcommand\een{\end{enumerate}}

\def\dfrac#1#2{{\displaystyle\frac{#1}{#2}}}
\newcommand\etal{{\hbox{{\textit\ et. al.\/}\textit\ }}}

\newcommand{\nunit}{\hbox{ cm}^{-3}}%
\newcommand{\rhounit}{\hbox{ g}\hbox{ cm}^{-3}}
\newcommand{\gev}{\hbox{ GeV}}
\newcommand{\ev}{\hbox{ eV}}
\newcommand{\mev}{\hbox{ MeV}}

\newcommand{\cm}{\hbox{ cm}}

\newcommand\hm{\mathbb H}

\newcommand\poi{Poincar$\acute{\rm e}$~}

\newcommand\schd{Schr$\ddot{\rm o}$dinger}

\newcommand\sch{Schr$\ddot{\rm o}$dinger~}

\newcommand\bloch{Bl$\ddot{\rm o}$ch~}

\newcommand\mobius{M$\ddot{\rm o}$bius~}

\def\ket#1{| \,#1\, \rangle}
\def\bra#1{\langle \,#1\, |}
\def\scx#1#2{\langle \,#1\, |\, #2\, \rangle}

\def\me#1#2#3{\langle \,#2\, |\,#1\,|\, #3\,\rangle}
\newif\ifboo \boofalse

\begin{document}
\textheight=23.8cm
\title{\Large Topological phase in two flavor neutrino oscillations}
\author{\bf Poonam Mehta}
\email{poonam@rri.res.in} \affiliation{Raman Research Institute, C.
V. Raman Avenue, Bangalore 560 080, India}
\date{\today}
%
\begin{abstract}
 We show that the phase appearing in  neutrino flavor oscillation formulae has a geometric
and topological contribution. We identify a topological phase
appearing in the two flavor neutrino oscillation formula using
Pancharatnam's prescription of quantum collapses between
nonorthogonal states. Such quantum collapses appear naturally  in
the expression for appearance and survival probabilities of
neutrinos. Our analysis applies to neutrinos propagating in vacuum
or through matter. For the minimal case of two flavors with $CP$
conservation, our study shows for the first time that there is a
geometric interpretation of the neutrino oscillation formulae for
the detection probability of neutrino species.
\end{abstract}
%
\maketitle  \vskip .6in
\section{Introduction}
 \label{intro}
 The phenomenon of neutrino flavor
oscillation results from the phase difference acquired by the mass
eigenstates due to their time evolution while propagating in vacuum
or in matter. The observation of neutrino flavor oscillations in
solar, atmospheric, reactor, and accelerator experiments reveal the
remarkable fact that the neutrinos exhibit sustained quantum
coherence even over astrophysical length
scales~\cite{conchamaltoni,funchal}. It is then natural to ask what
we can learn about neutrinos from
 these coherent phases. Here, we address the issue of geometric and
 topological phases involved in the physics of neutrino oscillations.

On the theoretical front, it is well known that the phenomenon of
neutrino oscillations cannot be accommodated within the standard
model (SM) of particle physics. Therefore, the experimental
observation of neutrino oscillations provides a concrete evidence
for the requirement of physics beyond the SM and neutrinos have been
an intensive area of research in the past several years.

 The study of geometric phases in the context of neutrino oscillations
 has been carried out in the past by several
 authors~\cite{nakagawa,vidal,as1,as,smirnovgeometric,smirnovgeometric2,akhmedovspin,guzzo,aquino,Naumov:1991,Naumov:1991rh,Naumov:1992,Naumov:1993vz,he,Blasone:1999tq,wang},
but none of the papers seem to provide a unified perspective on the
problem taking into account the different {\it avatars} of geometric
phase.  It is worthwhile to stress here that one needs to be
cautious while interpreting claims in the literature as they
crucially depend on which version of the geometric phase one is
dealing with. We will first summarize the related literature and
then focus on the specific question that we address in this paper.
We mostly restrict our attention to the case of two neutrino flavors
and the $CP$ ($CP$ stands for charge-conjugation and parity)
conserving situation, which is the minimal scenario for studying the
physics of oscillations. We find, in contrast to earlier studies of
this problem that the geometric phase appears even in this minimal
context.

Let us first review the papers that are connected to
Berry's~\cite{Berry:1984jv} cyclic adiabatic phase. Berry studied
phases that appear when the Hamiltonian of a quantum system depends
on parameters that are varied slowly and cyclically.
Nakagawa~\cite{nakagawa} followed this work by an elegant paper in
which he  pointed out that the geometric phase could also arise in
systems where adiabatic theorem did not hold. The key point made by
Nakagawa was that  while for existence of geometric phases,
adiabatic condition was not necessary (this was also independently
pointed out by Aharonov and Anandan~\cite{Aharonov:1987gg}), the
adiabatic theorem itself could be most easily understood in terms of
geometric arguments. As an application of his general formalism,
Nakagawa considered two flavor neutrino oscillations in matter. He
concluded that the Berry phase played no role in this situation. The
topological phase in the two flavor neutrino case, which is the
central result of the present paper was missed in his work because
he restricted himself to a limited region in the parameter (ray)
space and did not consider generalizations of the geometric phase
that allow for quantum collapse.

Subsequent work on Berry's geometric phase and neutrinos exploited
the spin degree of freedom of neutrinos and its interaction  with
the transverse  magnetic field leading to geometric effects and spin
flip. Since at that time, spin precession was a plausible solution
to the solar-neutrino problem, there is a body of work by several
authors on the subject of geometric phase effects in this context,
both in the absence and presence of matter and mass-splitting
terms~~\cite{vidal,as1,as,smirnovgeometric,smirnovgeometric2,akhmedovspin,guzzo,aquino}.
 However, in the present scenario, spin flavor precession is
disfavored as the leading solution to the solar-neutrino problem at
99.86 \% C.L.~\cite{miranda}, which makes it phenomenologically
uninteresting. Also, we would like to mention that in the present
study, spin plays only a passive role, and we shall not discuss this
particular aspect any further.

Naumov~\cite{Naumov:1991,Naumov:1991rh,Naumov:1992,Naumov:1993vz}
studied geometric phases for two and three flavor neutrino
oscillations taking into account the  optic
potentials~\cite{raffeltbook} induced by coherent forward scattering
of neutrinos against the background matter via SM interactions. The
slowly changing parameters in the Hamiltonian were identified as a
set of optic potentials $q(t)$, which were connected to the
refractive indices of neutrinos in a medium. For the naturally
existing cyclic cases like spherically symmetric or sandwich-like
density profiles, he found that the geometric (or topological) phase
was zero for both two and three flavors due to only one of the optic
potentials appearing in an {\it essential} manner in the
Hamiltonian. Note that the two terms ``topological" and ``geometric"
were used interchangeably in Naumov's works. Here we will make a
distinction between the two terms. The topological phase refers to
phase factors that are insensitive to small changes in the circuit,
while geometric phases are sensitive to such changes.

In a more recent paper, He \etal~\cite{he} carried out a detailed
study of the Berry phase in  neutrino oscillations for both two and
three flavors, active and sterile mixing, and with inclusion of
nonstandard interactions. For the particular case of two flavor
oscillations in matter, they claimed that the Berry phase can only
appear if  nonstandard (R-parity violating supersymmetry)
neutrino-matter interactions are taken into account.

All the above
papers~\cite{nakagawa,Naumov:1991,Naumov:1991rh,Naumov:1992,Naumov:1993vz,he}
claim that the geometric phases do not appear in the oscillation
probabilities for the case of two flavor neutrinos with $CP$
conservation in vacuum or in matter as long as neutrino-matter
interactions are standard {\it i.e.} coherent forward scattering is
induced by charged current interaction of electron neutrino
($\nu_e$) with electrons in matter. The above claims can be
understood as the necessity of having at least two {\it{essential}}
parameters in the Hamiltonian to detect curvature. Because of the
absence of flavor changing neutral currents
 in the SM, it turns out that for the case of ordinary electrically neutral matter,
even though one has two varying parameters - electron number density
($n_e$) and neutron  number density ($n_n$), only one of these will
appear in an essential way in the Hamiltonian and hence the Berry's
geometric phase is expected to be zero. The other parameter $n_n$
just adds a global phase to the time-evolved neutrino flavor state
and hence does not affect oscillation. But also it is worth
stressing that if both the conditions of having a nontrivial
multidimensional parameter space as well as cyclic evolution of the
states in parameter space were satisfied, the net geometric phase
(resulting from the difference between the geometric phases picked
up by the individual mass eigenstates) would have  appeared in the
formulae for detection probability and hence been observable.

Next, we will briefly review and summarize papers dealing with
geometric phases that are generalizations of the Berry
phase~\cite{Pancharatnam:1956,Aharonov:1987gg,sam} in the context of
neutrinos~\cite{Blasone:1999tq,wang}. Such geometric effects can
appear under less restrictive conditions than those required for
Berry's version of the geometric phase. Infact such phases can
appear even in situations where there are  no  parameters varying in
the Hamiltonian and the evolution is not necessarily cyclic or
unitary. Note, however, that in general the geometric phases
appearing in transition amplitudes are global phases that do not
have any observable consequences. To observe such a phase one needs
a split-beam interference experiment in which a beam is spatially
separated into two parts that suffer different histories. Such an
experiment is hard to design for neutrinos because they interact so
weakly and are nearly impossible to deflect or confine. This renders
such phases uninteresting as they are not observable as far as
neutrinos are concerned. Our aim  here is to explore whether there
are geometric effects that survive at the level of detection
probabilities that are directly measurable quantities.

 Blasone et. al.~\cite{Blasone:1999tq} claimed that Berry's phase
was present in the physics of neutrino oscillation in vacuum even
for the two flavor $CP$ conserving case. Their argument is based on
the fact that under \sch evolution, the pure flavor states come back
to themselves after one period ($T$) of oscillation having acquired
an overall phase. This overall phase was shown to be a sum of a pure
dynamical phase and a part that depended on the mixing angle only
and independent of energy and masses of the two mass states (hence,
geometric). They called this extra phase the Berry phase. Note that
this phase picked up by a neutrino flavor state arises purely due to
\sch evolution of the system giving a closed loop in the Hilbert
space but not due to any slowly varying parameters leading to
adiabatic evolution of the  Hamiltonian itself. Hence strictly
speaking it is the Aharonov-Anandan cyclic
phase~\cite{Aharonov:1987gg} that generalized Berry's adiabatic
phase to situations where the adiabaticity constraint did not apply
and only the cyclic condition is met. Also, we should note that
since the phase obtained was a global phase at the amplitude level,
it does not appear in measurable quantities like neutrino appearance
or survival probabilities as mentioned above.

After Berry's~\cite{Berry:1984jv} seminal paper on this subject,
Ramaseshan and Nityananda~\cite{ramaseshan} pointed out that Berry's
phase had a connection with the phase obtained by
Pancharatnam~\cite{Pancharatnam:1956} in the fifties in his study on
interference of polarized light. These insights were carried over to
the ray space of quantum mechanics by Samuel and
Bhandari~\cite{sam}. They showed that the two seemingly different
geometric phases obtained by Berry and Pancharatnam (appearing under
different  sets of conditions) could be described in a unified
framework. They also pointed out that geometric phases are not
restricted to unitary, cyclic and adiabatic
evolution~\cite{Berry:1984jv} of a quantum system and can appear in
an even more general context that allows for quantum collapses,
which occur during measurements. Following this line of thought,
Wang \etal~\cite{wang} extended the study of Blasone
\etal~\cite{Blasone:1999tq} to obtain noncyclic geometric phases for
two and three flavor neutrinos in vacuum. Their claim can be
understood as follows. Consider the \sch evolution of a quantum
state over an arbitrary time period from $\tau=0$ to $\tau$. Now
this open loop (noncyclic) \sch evolution of a quantum state over a
time $\tau$ can  be closed by a collapse of the time-evolved quantum
state at $\tau$ onto the original state at $\tau=0$ by the shorter
geodesic curve joining the two states in the ray space~\cite{sam}.
The phase associated with the complex number ($r e^{i\ss}$)
representing  the inner product of the original state vector and the
time-evolved state vector (with the dynamical phase removed) has a
pure geometric origin. This noncyclic  geometric phase  was
evaluated by Wang \etal~\cite{wang} for both the two and three
flavor cases. But, again note that this phase will be unobservable
as it only appears at the level of amplitude.

The main purpose of the present work is to establish that
Pancharatnam's phase  does appear in detection probabilities and
hence is directly observable. For the simplest case of two flavors
in vacuum or in constant density matter (restricting to SM
interactions) with $CP$ conservation, we obtain a Pancharatnam phase
of $\pi$, and this leads to an elegant geometric interpretation of
the neutrino oscillation formulae. We also make a direct connection
of this  phase with the Herzberg and Longuet-Higgins topological
phase~\cite{hlh} in molecular physics. We show that the Pancharatnam
phase of $\pi$ remains  even in the presence of slowly varying
matter density and this can be ascribed to the topological nature of
this phase. Inclusion of $CP$ violation can
 change the topological nature of the phase and make it a path-dependent geometric
phase.

Although one should do a full three flavor analysis for a complete
treatment, we work in an effective two flavor approximation that is
fairly justified~\cite{kuo,Akhmedov:2004ve} due to the smallness of
$\Theta_{13}$, and hierarchy of mass splittings ($\vert \delta
m_{21}^2/\delta m^2_{32} \vert << 1$) and in addition on matter
interactions being standard~\footnote{It turns out that in the
presence of non-standard interactions during propagation, it is
possible to do the analysis with only two flavors for the case of
solar neutrinos while a complete three flavor analysis is needed for
the case of the atmospheric neutrinos~\cite{iss}. }. In many
physical situations, observations depend on mainly one mixing and
one mass squared splitting. Conventionally, $\Theta_{12}$ and
$\delta m^2_{21}$ describe oscillations of solar neutrinos, while
$\Theta_{23}$ and $\delta m^2_{32}$ are used to describe atmospheric
neutrinos. The mixing angle $\Theta_{13}$ gives small effects on
both solar and atmospheric neutrinos. Working with only two flavors
is of course advantageous as the results obtained are physically
more transparent and can be visualized in analogous situations in
optics and the \poi sphere can be used as a calculational tool to
study the system.

For the ease of visualization of the phenomena of oscillations, in
the past several authors have discussed simple pictorial depiction
of neutrino oscillations in terms of precession of a (pseudo) spin
vector in three-dimensional space in a variety of contexts for the
case of two neutrino
flavors~\cite{harris,PhysRevD.35.4014,stodolsky,PhysRevD.37.1072,thomson,Enqvist1991754,giunti,kim}.
Below we give a brief account of the papers dealing with geometric
representation of  neutrino flavor oscillations.  Harris and
Stodolsky~\cite{harris}  addressed the question of a unified
treatment of generic two-state systems (including particle mixing
involving two neutrino types) in media using density matrices. It
was shown that the equation of motion for the polarization vector
represented the precession of polarization vector about a vector
representing an effective magnetic field (which could result from
the mass terms in vacuum or matter terms). Kim
\etal~\cite{PhysRevD.35.4014} discussed the analogy of
solar-neutrino oscillations with that of precession of electron spin
in a time-dependent magnetic field. They applied this picture in the
limit of adiabatic approximation.
Stodolsky~\cite{stodolsky} described the evolution of a statistical
ensemble (neutrinos from supernovae or in the early Universe)
applying the density matrix approach~\cite{harris} and showed that
oscillations in presence of mixing and matter interactions in a
thermal environment could be viewed in terms of precession. Kim
\etal~\cite{PhysRevD.37.1072} derived the geometric picture for two
and three flavor neutrinos and applied it to nonadiabatic
 as well as adiabatic cases. Thomson and McKellar~\cite{thomson}
 treated the case of neutrino background giving rise to nonlinear
 feedback terms in the equation of motion  for polarization vectors
 and gave a pictorial representation for the same.
Enqvist \etal~\cite{Enqvist1991754}  describe visualization of
oscillations of a thermal neutrino ensemble of the early Universe.
The geometrical representation in wave packet treatment of
oscillations was discussed by Giunti \etal~\cite{giunti}. As in
optics, the \poi sphere is a convenient tool for visualizations and
calculations pertaining to neutrino oscillations, particularly in
looking for geometric effects.

 This paper is organized as follows. In
Sec.~\ref{analogy}, we develop an analogy between the neutrino
flavor states and polarized states in optics since such a mapping
allows for a convenient visualization of geometric effects. We then
go on to show in Sec.~\ref{panc} that the Pancharatnam phase does
appear in the detection probabilities of neutrino species in the two
flavor neutrino system in vacuum and also in matter. We conclude
with a discussion of our key result and future directions in
Sec.~\ref{disc}. Throughout we set $\hbar=c=1$.

 \vskip .6 true cm
\section{Correspondence between two flavor neutrinos and polarization states in optics}
 \label{analogy}

Since the concept of Pancharatnam's phase was developed in the
 context of optics, it is worthwhile to first develop a
 correspondence between the mathematics of two flavor neutrino
 states and polarization states in optics.
 Let us first recall the conditions under which
 the two flavor neutrinos and polarization states in optics can be
 analyzed within an unified framework.

 \vskip .6 true cm
\subsection{Two flavor neutrinos}
 In the ultra-relativistic limit, the Dirac equation for two flavor
 neutrinos (antineutrinos) can be reduced to a
\sch form~\cite{raffeltbook,halprin} written in terms of a
two-component vector of positive (negative) energy probability
amplitude. This is analogous to Maxwell's equations reducing to the
linear \sch form for the polarization states in optics in the
paraxial limit~\cite{mukundaparaxial}.

The two neutrino  flavor states can be mapped to a two-level quantum
system with distinct energy eigenvalues, $E_i \simeq p + {m_i^2}/{2
 p}$ in the ultrarelativistic
limit along with the assumption of equal fixed momenta (or
energy)~\cite{yossi,kim}. In the presence of matter, the
relativistic dispersion relation $E_i = f (p, m_i)$ gets modified
due to the neutrino-matter interactions (in an electrically neutral
homogeneous medium) leading to
\bea E_{i=\mp} &=& \left(
p + \dfrac{m_1^2 + m_2^2}{4p} + \dfrac{V_C}{2} + V_N \right) \nonumber\\
&& \mp \dfrac{1}{2} \sqrt{\left({\omega} \sin 2 \Theta\right)^2 +
\left( V_C - {\omega} \cos 2 \Theta  \right)^ 2}~, \label{eqone}
\eea
where $\omega=\delta m^2/2p$ with mass splitting $\delta m^2 =
m_2^2-m_1^2$ and
 $p \simeq E$ being the fixed momentum (energy) of the neutrino.
$\Theta$ is the mixing angle in vacuum.
 $V_C=\sqrt{2} G_F n_e = 7.6 \times 10^{-14} Y_e \rho \ev$ and
$ V_N=-\sqrt{2} G_F n_n/2 = - 3.8 \times 10^{-14} Y_n \rho \ev$  are
the respective effective potentials due to coherent forward
scattering of neutrinos with electrons (via charged current
interactions) and neutrons (via neutral current interactions).
$G_F=1.16637 \times 10^{-5} \gev^{-2}$ parameterizes  the  weak
interaction strength (Fermi constant). $V_C$ and $V_N$ depend on the
electron ($n_e$) and neutron ($n_n$) number densities (in units of
$\nunit$). $n_{e/n} = \rho Y_{e/n} N_{Avo}$, where $\rho $ is the
mass density in $\rhounit$, $Y_{e/n}$ is the relative electron
(neutron) number density and its value is roughly $\sim 0.5$ for
Earth matter, and $N_{{Avo}}$ is the Avogadro's number. Setting
$V_C=V_N=0$, we recover the vacuum case.

 Note the fact that although there are two densities
$n_e$ and $n_n$ appearing in the eigenvalues, it is only $n_e$ that
appears in a nontrivial way (through $V_C$) in the flavor
Hamiltonian,
 \bea {\mathbb
 H}_{\nu}^{\mathrm{}} &=&   \left( p + \dfrac{m_1^2+m_2^2}{4p} +
\dfrac{V_C}{2}+V_N
 \right)  \mathbb{I} \nonumber\\
 &+& \dfrac{1}{2} \left(
\begin{array}{cc}
 V_C -  \omega \cos  2\Theta  & \omega \sin  2\Theta \\
   \omega \sin  2\Theta  &  - (V_C - \omega \cos  2\Theta)
\end{array}
\right)~. \label{nuham} \eea The above Hamiltonian (Eq.~\ref{nuham})
also describes an inhomogeneous medium provided the  scale of
variation of matter induced potential $V_C$  is slow compared to the
 scale of the order of $\hbar/(E_+ - E_-)$), hence ensuring no
transitions between the mass eigenstates. This defines the
adiabaticity condition~\cite{yossi,kim}. As neutrinos traverse a
density gradient, at a particular value of  $n_e$ the diagonal
elements of ${\mathbb H}_{\nu}^{\mathrm{}}$ can vanish causing an
interchange of flavors irrespective of the value of the vacuum
mixing angle $\Theta$. This phenomenon of resonant conversion in
matter is known as the Mikheyev-Smirnov-Wolfenstein (MSW)
effect~\cite{Wolfenstein:1977ue,ms}.

The off-diagonal form of the Hamiltonian in  flavor basis (both in
vacuum and matter) leads to flavor oscillations of neutrinos, which
is the only mechanism that mixes the neutrinos of different
generations or flavors while preserving the lepton number (note that
the absence of flavor changing neutral currents prevents any flavor
change within the SM). Also note that the matter term appears in
diagonal elements only so in the absence of vacuum mixing, neutrinos
of different flavors cannot mix.
 The term proportional to  the identity gives an overall phase
to each of the mass eigenstates and hence does not affect
oscillations. This corresponds to the gauge freedom of any state of
a two-level quantum system~\cite{nakagawa}.

In the next subsection, we describe the polarized states in optics
in the language of quantum mechanics.

 \vskip .6 true cm
\subsection{Polarized states in optics}

Polarization optics is  mathematically identical to the evolution of
a two state quantum system. In a helicity basis for polarized light,
we can write $\ket{\mathfrak {R}} $ and $ \ket{\mathfrak {L}}$
representing right and left circular polarizations. A general
polarized light beam $\ket{\Psi} $ can then be expanded in this
basis as $\ket{\Psi} = \al \ket{\mathfrak {R}} + \be \ket{\mathfrak
{L}}$ where $|\al|^2 +|\be|^2 = N$, the intensity of the beam of
polarized light. We can parameterize an arbitrary state of polarized
light by \beq \ket{\Psi} = \sqrt{N} \exp\{i \eta \}\left(
\begin{array}{cc}
   \cos  (\theta/2)  \exp(-i \phi/2)  \\
   \sin  (\theta/2) \exp(i \phi/2)
\end{array}
\right)~, \label{state} \eeq where $N$ is the total intensity, which
is normalized to unity, and the angles $\theta$ and $\phi$ (where $0
\leq \theta \leq \pi$ and $0 \leq \phi \leq 2\pi$) describe the
state of polarization of the beam, represented on the
two-dimensional unit  sphere ($\mathbb S^2$) called the \poi sphere.
Orthogonal polarization states are antipodal points of the sphere.
$\eta$ is the overall phase of the beam. The states on the sphere
are defined modulo this overall phase of $\eta$ and represent the
ray space~\cite{samray}. The north pole ($\theta = 0$) represents
right circular light and the south pole ($\theta=\pi$) represents
left circular light. States on the equator ($\theta=\pi/2$)
represent linear polarizations. Any other point on the surface of
the sphere represents elliptic polarization. The \poi sphere is a
useful device to visualize the changes in the state of polarization
of a light beam traversing through a medium.

The mapping between the polarized states and a two-level quantum
system originates from the following fact. Neglecting absorption
effects~\footnote{The incoherent scattering cross section for
neutrinos ($10^{-44} \cm^2$ for 1 $\mev$ neutrinos impinging on
target of mass 1 $\mev$) is extremely small as compared to photons
in a medium. }, the effect of different media can be encoded in
terms of $2 \times 2 $ Hermitian matrix (Hamiltonian). The time
evolution of optical states in a medium is governed by a \schd-like
equation with the medium represented by the most general form of
Hamiltonian for a two-level system given by \bea
\mathbb{H} &=& {\mathcal A} \sigma_x + {\mathcal B} \sigma_y +
{\mathcal C} \sigma_z + {\mathcal D}\mathbb{I}~, \label{hmed} \eea
where, the coefficients of the three traceless Pauli matrices,
${\mathcal{A, B}}$  and ${\mathcal{C}}$  are responsible for
generating rotations of incident optical states about
 $x,y,z$ axes on the \poi sphere.
  $\mathcal D$ just adds an overall phase that can be absorbed in a
redefinition of the state.
%
%
\begin{figure}[htb]
\centering \vspace*{3mm} \hspace*{1mm}
\epsfig{figure=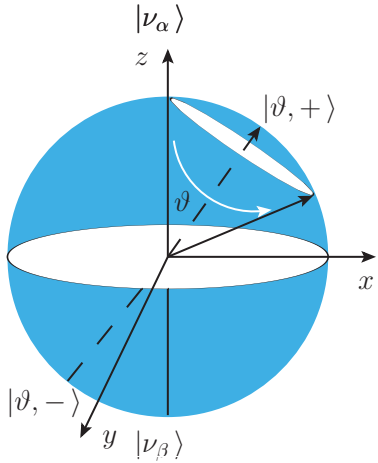,width =0.7\columnwidth} \vspace*{-1mm}
\caption{Neutrino states on the \poi sphere. The flavor states
$\ket{\nu_\al}$ and $\ket{\nu_\be}$ are the two antipodal points on
the $z$ axis while $\ket{\vartheta,{\pm}}$ correspond to the mass
(energy) eigenstates lying on an axis making an angle $\vartheta$
with respect to the $z$ axis. } \label{fig1}\end{figure}
Hence given an arbitrary medium, it can be represented by a
Hamiltonian as mentioned above, and the eigenstates of the
Hamiltonian represent those optical states that do not suffer any
change (when incident on such a medium) in their state of
polarization except for picking up an overall phase shift. The
polarization of any other state (other than the eigenstates)
incident on this medium will undergo a periodic change. On the \poi
sphere this can be visualized as a rotation of the incident state
vector about the axis defined by a line joining the two eigenstates
of the Hamiltonian.
Mathematically, these  unitary rotations on the \poi sphere are
generated by $e^{-i \mathbb{H} t}$. This is identical to unitary
time evolution generated by the Hamiltonian of the quantum states in
the Hilbert space.  The quantum-mechanical analogue of the \poi
sphere is the \bloch
  sphere, which geometrically represents the space of pure states of a
  two-level quantum system.

Nonvanishing values of ${\mathcal{A, B, C}}$ simultaneously
parameterize the effect of an elliptically birefringent medium.
 Circular (linear) birefringence are
 special cases where the conditions ${\mathcal{A, B}}=0$ and ${\mathcal{C, D}} \neq 0$
 (${\mathcal{B, C}}=0$ and ${\mathcal{A, D}} \neq 0$) are
satisfied.

 \vskip .6 true cm
\subsection{Neutrinos and  optics analogy}

We can now describe the isomorphism between neutrino states and
polarized states in optics. The complete set of states for two
flavor neutrino system can be represented on the \poi sphere just
like the optical states as depicted in Fig~\ref{fig1}. For
convenience we define a new coordinate $\vartheta$, which goes from
$ 0 \to 2\pi$ as we traverse the unit great circle in the $x-z $
plane. In terms of the old coordinates, the points $\theta,\phi=0$
are now labeled by $\vartheta=\theta$ and the points
$\theta,\phi=\pi$ are labeled by $\vartheta=2\pi-\theta$.
If we assume that the flavor states are the north and south poles of
the \poi sphere, then the mass eigenstates are represented by the
two antipodal points lying on an axis making an angle $2 \Theta =
\vartheta$ with respect to the polar axis. States on the equator
   coincide with the mass eigenstates for the special case of
   maximal mixing ($\Theta=\vartheta/2=\pi/4$) which
   corresponds to complete flavor conversion (MSW effect).
   Geometrically, the MSW effect can be viewed as rotation about an
   equatorial axis, rotating the north pole into the south pole.

Ignoring the term proportional to the Identity, the neutrino
Hamiltonian (Eq.~\ref{nuham}) both in vacuum or matter can be recast
in exactly the same form given by (see Eq.~\ref{hmed})
 \bea
 \hm_{\nu}
 &=&  \dfrac{\omega}{2} \left[ (\sin \vartheta) \sigma_x -(\cos  \vartheta) \sigma_z
 \right]~,
\label{nuham2}\eea where $\omega=\delta m^2/2p$ and the mixing angle
$\Theta$ is replaced by $\vartheta/2$~\footnote{In defining the \poi
sphere,
   it is useful to work with half angles $\vartheta/2$ as it allows for a mapping of
   the entire set of states on to a two-dimensional sphere $\mathbb{S}_2$ as
   $\vartheta$ changes from $0 $ to $4\pi$.}.
Comparing  the two Hamiltonians (Eq.~\ref{hmed} and
Eq.~\ref{nuham2}) we see that the  neutrino Hamiltonian represents a
medium with elliptic birefringence. And neutrino oscillations can be
viewed as the neutrino flavor state precessing~\cite{kim} about the
line joining the mass eigenstates (analogous to elliptic axis)
induced by the time-evolution operator $e^{-i\mathbb{H}_\nu t}$ on
the \poi sphere. In the language of neutrino optics, both vacuum and
matter exhibit elliptic birefringence property with different
elliptic axes.

The absence of flavor changing neutral currents in the SM gives rise
to a real form of the Hamiltonian ($\mathcal{B}=0$), and it
corresponds to a $CP$-conserving situation. The eigenvectors (also
called mass eigenstates) of Eq.~\ref{nuham2} are given by \beq
\ket{\vartheta,+} =
\begin{pmatrix}
 \cos (\vartheta/2)   \\
 \sin (\vartheta/2)
 \end{pmatrix} ~{\mathrm{and}}~ \ket{\vartheta,-} = \begin{pmatrix}
 -\sin (\vartheta/2)   \\
 \cos (\vartheta/2)
 \end{pmatrix}~. \label{evec}
\eeq Note that states $\ket{\vartheta,+}$ and $ \ket{\vartheta,-}$
are orthogonal antipodal points
 on the \poi sphere which always lie on the great circle formed by
the intersection of the
 $x-z$ plane with the \poi sphere.
 Mass eigenstates lying outside the $x-z$ plane
  imply  $CP$ violation. This fact has very interesting consequences
 for the physics of geometric phases in $CP$ nonconserving situations~\cite{pmcpv}.

\vskip .6 true cm
\section{Pancharatnam's phase in the two flavor neutrino system}
\label{panc}

{\it{The Pancharatnam phase :-}} We give a  brief introduction to
the idea of Pancharatnam's phase in quantum-mechanical language
along the lines of Ref.~\cite{samray,berrypancharatnam,sam}. Given
any two nonorthogonal states $\ket{\mathfrak{A}}$ and
$\ket{\mathfrak {B}}$ in the Hilbert space describing a system, a
notion of geometric parallelism between the two states can be drawn
from the inner product $\scx{\mathfrak{A}}{\mathfrak{B}}$. The two
states are said to be parallel (in phase) if
$\scx{\mathfrak{A}}{\mathfrak{B}}$ is real and positive, which
defines  {\it{the Pancharatnam connection (or rule)}}.
Geometrically, it implies that the norm of the vector sum of the two
states $||(\ket{\mathfrak{A}} + \ket{\mathfrak{B}})||^2 =
\scx{\mathfrak{A}}{\mathfrak{A}} + \scx{\mathfrak{B}}{\mathfrak{B}}
+ 2 |\scx{\mathfrak{A}}{\mathfrak{B}}| \cos ({\rm {ph}}
\scx{\mathfrak{A}}{\mathfrak{B}}) $ is maximum. Physically, it
implies that if we let the two states  interfere with each other the
resulting state will have maximum probability (intensity). Note that
if $\ket{\mathfrak A}$ is in phase with $\ket{\mathfrak B}$, and
$\ket{\mathfrak B}$ is in phase with $\ket{\mathfrak C}$, then
$\ket{\mathfrak C}$ is not necessarily in phase with state
$\ket{\mathfrak A}$. The  phase difference between the  states
$\ket{\mathfrak C}$  and $\ket{\mathfrak A}$ is {\it{the
Pancharatnam phase}}, and it is equal to half the solid angle
$\Omega$ subtended by the geodesic triangle $\mathfrak{A,B,C}$ on
the \poi sphere for a two-level system at its center. In general,
for an $n$-level system, the space of states is given by
 ${\mathcal{CP}}^{n-1}$ ($\mathcal{CP}$ stands for complex projective)
 which reduces to the \poi sphere ($\mathbb{S}^2$) for a two-level system ($n=2$).
Nonintegrability of Pancharatnam's connection follows from the
nontransitivity of the rule.

Pancharatnam's phase reflects the curvature of projective Hilbert
space (ray space)
 and is independent of any parameterization or slow variation.
Thus it can also appear in situations where the Hamiltonian is
constant in time.
All one needs is that the state has a nontrivial trajectory on the
\poi sphere. This condition is met naturally for neutrinos since
they are produced and detected as flavor states (which are not the
stationary mass eigenstates) and hence they automatically explore
the curvature of the ray space (\poi sphere) under the \sch time
evolution.
Furthermore, note the fact that \sch evolution (possibly)
interrupted by measurements can lead to Pancharatnam's phase. If we
take any state and subject it to multiple quantum collapses (such
that consecutive collapses are between nonorthogonal states) and
bring it back to itself, then the resulting state is given by
$\ket{\mathfrak A} \scx{\mathfrak A}{\mathfrak C} \scx{\mathfrak
C}{\mathfrak B} \scx{\mathfrak B}{\mathfrak A} $, where the phase of
the complex number $\scx{\mathfrak A}{\mathfrak C} \scx{\mathfrak
C}{\mathfrak B} \scx{\mathfrak B}{\mathfrak A} $ is given by
$\Omega/2$.

{\it{The Herzberg and Longuet-Higgins phase and $CP$-conserving
neutrino Hamiltonian :- }} Let us reexamine the form of the neutrino
Hamiltonian given by Eq.~\ref{nuham2} and the eigenvectors given by
Eq.~\ref{evec}. Note that the eigenvectors depend only on a single
parameter $\vartheta $ and satisfy
\bea
\ket{\vartheta,\pm} &=& \mp \ket{\vartheta+\pi,\mp} =
-\ket{\vartheta+2\pi,\pm}  \nonumber\\
 &=& \pm \ket{\vartheta+3\pi,\mp}= \ket{\vartheta+4\pi,\pm}~. \eea
  The minus sign picked up by both the
mass eigenstates as we change $\vartheta$ from $0 \to 2 \pi$ is
precisely {\it {the Herzberg and Longuet-Higgins
phase}}~\cite{hlh,shaperebook} of $\pi$, which was first obtained in
the context of molecular physics in 1963. So, we note that just by
looking at the form of the Hamiltonian for neutrino system, we
should expect the  Herzberg and Longuet-Higgins phase to appear.
Also, note that the space of rays for the real neutrino Hamiltonian
is the great circle ($\mathbb{S}^1$) lying on the $x-z$ plane of the
\poi sphere (Fig.~\ref{fig1}) and  global structure of the
eigenvectors is a {\it{M{\" o}bius band}}.
The variation of $\vartheta$ results in parallel transport of the
mass eigenstates (with dynamical phase removed) following the
parallel transport rule along $\vartheta$ , \beq \Im m
{\me{\dfrac{d}{d\vartheta}}{\vartheta_\mp}{\vartheta_\mp}} = 0~.
\label{parallel} \eeq This parallel transport rule (formally
referred to as natural connection) has an anholonomy defined on the
\mobius band and  this leads to the topological phase of $\pi$. The
 topological phase factor ${\ss}$ depends on the vector potential
$A_{\vartheta}$ given by
\beq   {\ss} = \oint A_{\vartheta} d \vartheta = \oint \Im m
\me{\dfrac{d}{d\vartheta}}{\vartheta_\mp}{\vartheta_\mp} \,d
\vartheta~. \label{betadefn} \eeq This vector potential
$A_\vartheta$ is nonintegrable, and
 this is the  anholonomy of the connection.
Physically, this corresponds to  half a unit of  magnetic flux
 piercing the origin of the $x-z$ plane,
encircling which leads to this topological phase. And, the origin of
the circle is connected to the null Hamiltonian ({\it{i.e.}} all
elements are zero), which corresponds to the degeneracy point.

 Naively speaking, one would think that this phase will be impossible
 to access for neutrinos because we do not have a handle on the mixing angle $\vartheta/2$ to be varied
in a controlled way from $\vartheta=0 \to 2 \pi$. The key point to
understand here is the fact that as long as we carry out a quantum
evolution of a state in a closed loop enclosing the point of
singularity (degeneracy point, origin of the \poi sphere), which can
be achieved either via adiabatic variation of $\vartheta$ or via
\sch evolution interrupted by collapses, one will always  get this
phase. However, note that in the former case, the amplitude of the
initial state undergoing evolution does not change but in the latter
case, it diminishes. In what follows, we will show that the
transition probability for neutrinos actually does carry imprints of
such a topological phase, which can be explicitly derived using
Pancharatnam's prescription. We then show that the phase of $\pi$
actually appears there and is in fact observed by all the
experiments carried out so far.

{\it{The topological phase in two flavor neutrino oscillations
(invoking collapses and adiabatic evolution) :-}} In what follows,
we consider the most general situation, i.e. neutrinos are
traversing through matter with slowly varying density (i.e.
$\vartheta$ is a slowly varying parameter changing from
$\vartheta_1$ to $\vartheta_2$). Vacuum or constant density matter
will be special cases where $\vartheta$ is a constant.

\begin{figure}[htb]
\centering \vspace*{3mm} \hspace*{1mm}
\epsfig{figure=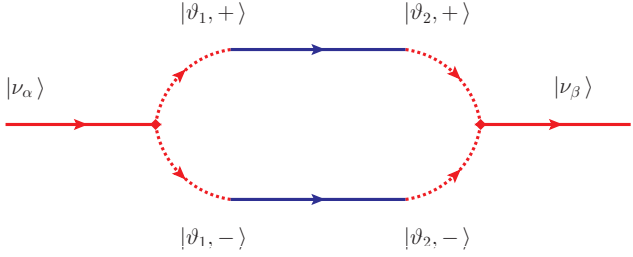,width =1\columnwidth} \vspace*{-1mm}
\caption{Schematic of a  split-beam experiment for neutrinos in
energy space. $\ket{\nu_\al}$ and $\ket{\nu_\be}$ are the two flavor
states, while $\ket{\vartheta_1,{\pm}}$ and also
$\ket{\vartheta_2,{\pm}}$ correspond to two sets of mass (energy)
eigenstates. $\ket{\vartheta_1,{\pm}}$ are adiabatically evolved
 to states $\ket{\vartheta_2,{\pm}}$, respectively (upon removing the dynamical phase). }
 \label{fig2}\end{figure}

 In order to see the effect of geometric phases,
usually one performs a split-beam experiment. In the case of optics,
one separates a beam into two parts in space and each part traverses
a different path. Finally the beams are recombined to observe the
relative phase shift as they interfere. In optics, the reflective
and  refractive property of the medium is exploited to make devices
like mirrors and lenses, which facilitates designing of such
experiments in the laboratory.
 In the case of neutrinos, such a procedure is not possible
owing to the fact that
 the refractive index is extremely small ($n_{refr}-1 \simeq 10^{-19}$ for
 neutrinos of energy $1 \mev$ in ordinary matter).
 Treating the Sun (with density $\rho=150 \rhounit$ in the core) as a
spherical lens for a neutrino beam of energy $10 \mev$ passing
through it, one gets the focal length to be  around $10^{18}
R_{\odot}$~\cite{raffeltbook}, which is about
 $10^5 $ times the size of our galaxy. Spatially split-beam
 interference experiments with neutrinos are clearly impossible.
However, the fact that neutrinos are produced and detected as flavor
states allows us to think of the time evolution of neutrinos as a
split-beam experiment in  energy space as depicted in
Fig.~\ref{fig2}.

 Let us consider a neutrino created as a flavor state $\ket{\nu_\al}$ (for example, neutrinos
 produced inside the Sun are predominantly in the electron neutrino flavor state, $\ket{\nu_e}$)
 and detected as another flavor state, $\ket{\nu_ \be}$ ($
 \ket{\nu_\be}$ can either be a $\ket{\nu_e}$, {\it i.e.} survival of the same electron neutrino flavor or a
 $\ket{\nu_\mu}$, {\it i.e.} appearance of muon neutrino flavor), then
\bea \ket{\nu_\al} = \nu_{\al +} \ket{\vartheta_1,{+}} + \nu_{\al -}
\ket{\vartheta_1,{-}}~, \eea where $\ket{\vartheta_1,{\pm}}$ are the
eigenstates of $\mathbb{H}_\nu (\vartheta_1)$. Now we consider an
adiabatic evolution of the mass eigenstates from
$\ket{\vartheta_1,{\pm}}$ to $\ket{\vartheta_2,{\pm}}$ due to a slow
enough variation of background density such that no mixing between
the two eigenstates is ensured under time evolution, and
$\ket{\vartheta_1,{\pm}}$ evolves to
\bea \ket{\vartheta_1,{\pm}} &\to&
 e^{-
i {\mathcal D}_\pm} \ket{\vartheta_2,{\pm}} \quad {\rm {with}}
\nonumber\\
{\mathcal D}_\pm &=&  \pm  \dfrac{1}{2} \int_0^t
\sqrt{\left({\omega} \sin \vartheta\right)^2 + \left( V_C - {\omega}
\cos \vartheta \right)^ 2} dt' \nonumber\\ &+& \int_0^t \left( p +
\dfrac{m_1^2 + m_2^2}{4p} + \dfrac{V_C}{2} + V_N \right) dt'~,
\label{correctadiab}\eea as the dynamical phases, relevant both for
the vacuum case ($V_C=V_N=0$) and in the presence of varying matter
density profile and $t$ is the time of flight of the neutrino. The
quantities  that depend on time (or distance) are $V_C$ and $V_N$
defined earlier (see Eq.~\ref{eqone}). Note that the states
$\ket{\vartheta_1,{\pm}}$ are $\ket{\vartheta_2,{\pm}}$ are
 connected via parallel  transport rule (Eq.~\ref{parallel}) on the \poi
 sphere.
 The two time-evolved states $e^{- i {\mathcal D}_\pm}
\ket{\vartheta_2,{\pm}}$ are finally recombined to form a flavor
state at the detector.
%
\begin{figure*}[t]
\centering \vspace*{-40mm} \hspace*{2mm}
\epsfig{figure=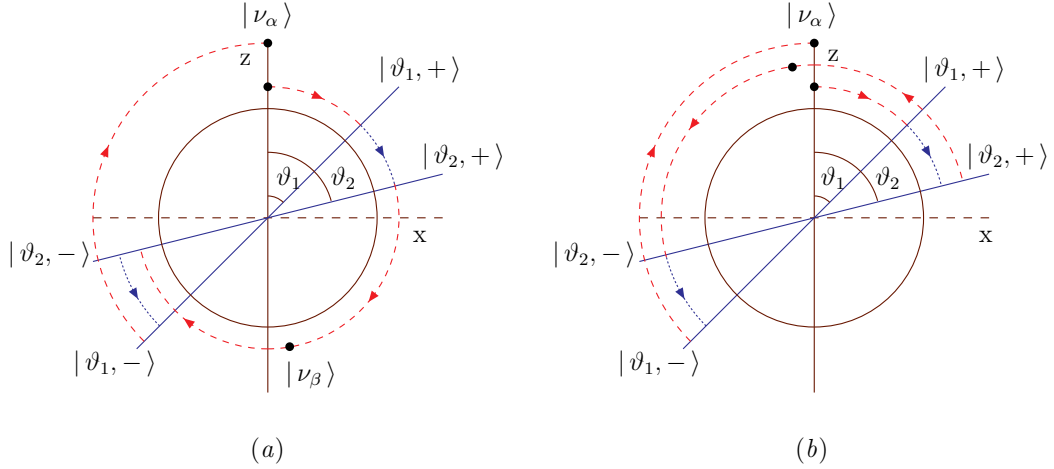,width =2\columnwidth} \vspace*{-15cm}
\caption{Two representative cases depicting the collapse processes
(dashed red lines) with intermediate adiabatic evolutions upon
removing the dynamical phase (dotted blue lines) on the great circle
($\mathbb{S}^1$) arising due to the
 cross term
$\scx{\nu_\al}{\vartheta_1,{-}} \scx{\vartheta _2,{-}}{\nu_\be}
\scx{\nu_\be}{\vartheta_2,{+}} \scx{\vartheta_1,{+}}{\nu_\al}$ in
the probability. The initial flavor state $\ket{\nu_\al}$ is on the
positive $z$ axis, while the final flavor state $\ket{\nu_\be}$ is
not necessarily its antipodal point. The two sets of mass
eigenstates are antipodal points on two axes making angles
$\vartheta_{1}$ and $\vartheta_{2}$ respectively with respect to the
$z$ axis. Case (a) corresponds to appearance probability [${\cal P}
({\nu_\al \to \nu_\be}$)] for which we get a cyclic loop in
$\vartheta$ space. (b) The collapse processes for survival
probability  [${\cal P} ({\nu_\al \to \nu_\al}) $] do not enclose
any loop.
 \label{fig3}}
\end{figure*}
%

In order to see this explicitly, let us proceed as follows: The
amplitude for the transition between states ${\nu_\al} \to
{\nu_\be}$ is given by \bea{\cal A} (\nu_\al \to \nu_\be) =
\me{\mathcal{U}}{\nu_\be}{\nu_\al}~, \eea where $\mathcal{U}$ is the
unitary evolution operator given by \bea \label{ueq} \mathcal{U} &=&
e^{-i{\mathcal D}_+ } \ket{\vartheta_2,{+}} \bra{\vartheta_1,{+}} +
e^{-i{\mathcal D}_- } \ket{\vartheta_2,{-}} \bra{\vartheta_1,{-}}~.
\nonumber\\ \eea
Inserting two complete sets of states in the amplitude,
 \bea \label{ampone} {\cal A} (\nu_\al \to \nu_\be)
&& \!\!\!\!\!\! = \sum_{i,j=-}^{+} \scx{\nu_\be}{\vartheta_2,{i}}
\me{\mathcal{U}}{\vartheta_2,{i}}{\vartheta_1,{j}}
\scx{\vartheta_1,{j}}{\nu_\al} \nonumber\\ &&
\!\!\!\!\!\!\!\!\!\!\!\!\!\!\!\!\!\!\!\!\!\!\!
 =~ \scx{\nu_\be}{\vartheta_2,{+}}
\me{\mathcal{U}}{\vartheta_2, {+}}{\vartheta_1,{+}}
\scx{\vartheta_1,{+}}{\nu_\al} \nonumber\\&&
\!\!\!\!\!\!\!\!\!\!\!\!\!\!\!\!\!\!\!\!\!\!\! +~
\scx{\nu_\be}{\vartheta_2,{-}}
\me{\mathcal{U}}{\vartheta_2,{-}}{\vartheta_1,{-}}
\scx{\vartheta_1,{-}}{\nu_\al}~.
  \eea
Note that the cross terms do not contribute in the adiabatic limit.
     Upon substituting Eq.~\ref{ueq} in Eq.~\ref{ampone}, we get \bea {\cal A}
(\nu_\al \to \nu_\be) &=& e^{ -i {\mathcal D}_{+}}
\scx{\nu_\be}{\vartheta_2,{+}}
\scx{\vartheta_1,{+}}{\nu_\al} \nonumber\\
&+& e^{ -i {\mathcal D}_{-}} \scx{\nu_\be}{\vartheta_2,{-}}
\scx{\vartheta_1,{-}}{\nu_\al}~. \label{amp}  \eea  Then the
probability for flavor transition ${\nu_\al} \to {\nu_\be}$ is given
by
\bea {\cal P} (\nu_\al \to \nu_\be) &=& |{\cal A} (\nu_\al \to
\nu_\be) |^2 \nonumber\\ && \!\!\!\!\!\!  \!\!\!\!\!\!
\!\!\!\!\!\!\!\!\!\!\!\! \!\!\!\!\!\!  =
\scx{\nu_\alpha}{\vartheta_1,{+}} \scx{\vartheta _2,{+}}{\nu_\be}
\scx{\nu_\beta}{\vartheta_2,{+}} \scx{\vartheta_1,{+}}{\nu_\al}
\nonumber\\ && \!\!\!\!\!\!  \!\!\!\!\!\! \!\!\!\!\!\!\!\!\!\!\!\!
\!\!\!\!\!\!
 +~ \scx{\nu_\al}{\vartheta_1,{-}} \scx{\vartheta
_2,{-}}{\nu_\be}
\scx{\nu_\be}{\vartheta_2,{-}} \scx{\vartheta_1,{-}}{\nu_\al} \nonumber \\
 &&
\!\!\!\!\!\!  \!\!\!\!\!\! \!\!\!\!\!\!\!\!\!\!\!\!\!\!\!\!\!\!
 +~  [\scx{\nu_\al}{\vartheta_1,{-}}e^{ i {{\cal D}_-}}  \scx{\vartheta
_2,{-}}{\nu_\be} \scx{\nu_\be}{\vartheta_2,{+}} e^{ -i {\cal D}_+}
\nonumber\\
&&  \!\!\!\!\!\!\!\!\!\!\!\!\!\!\!\!\!\!\!\!\!\!\!\!\!
\scx{\vartheta_1,{+}}{\nu_\al} + {\rm{c.c.}} ]~. \label{prob} \eea
 The cross term term in Eq.~\ref{prob} is related to the interference
 term resulting from the two path interferometer depicted in Fig.~\ref{fig2}.
 Upon dropping the
 dynamical phase, we have $\scx{\nu_\al}{\vartheta_1,{-}} \scx{\vartheta _2,{-}}{\nu_\be}
\scx{\nu_\be}{\vartheta_2,{+}} \scx{\vartheta_1,{+}}{\nu_\al}$ which
can be viewed as a series of closed loop quantum collapses with
intermediate adiabatic evolutions given by $\ket{\nu_\al} \to
\ket{\vartheta_1,{+}} \to \ket{\vartheta_2,{+}} \to \ket{\nu_\be}
\to \ket{\vartheta_2,{-}} \to \ket{\vartheta_1,{-}} \to
\ket{\nu_\al} $ that  essentially covers a great circle in the $x-z$
plane as is shown in Fig.~\ref{fig3}(a). This closed trajectory
subtends a solid angle of $\Omega=2\pi$ at the center of the great
circle.
Hence without any further calculation, we can immediately predict
that the phase of the interference term will be $\pi$ (half the
solid angle) due to Pancharatnam's prescription. On the circle, each
of the individual collapse processes which essentially projects a
state with given angle $\vartheta$ to another state with different
angle $\vartheta'$ can be thought of as an infinite series of
infinitesimally close collapses between states defined as
$\ket{\vartheta}$ and $\ket{\vartheta+\delta \vartheta}$ as far as
geometric phases are concerned. The entire closed loop of collapses
with intermediate adiabatic evolutions mentioned above can be viewed
as a smooth variation  of $\vartheta$ from $0 \to 2\pi$ in the limit
$\delta \vartheta \to 0$ hence making a direct connection to the
Herzberg and Longuet-Higgins phase mentioned above. Nonetheless, we
must note that the  evolution of a state is unitary under
infinitesimal collapses ($\delta \vartheta \to 0$ limit) while it is
nonunitary under finite collapses leading to a loss in intensity
(probability). But the geometric phase of the evolving state remains
unaltered for the two cases mentioned above.

For the case when $\alpha = \beta$, i.e. survival
 probability, it is easy to see that the collapses do not form a closed loop enclosing the origin
 and therefore the interference term will not pick up any phase. This case  is
 depicted in Fig.~\ref{fig3}(b).

In a simpler situation when $\vartheta$ does not change, {\it i.e.}
the case of vacuum or constant density matter, the number of states
will be fewer (in the absence of variation of density,
$\ket{\vartheta_1,{\pm}}$ is the same as $\ket{\vartheta_2,{ \pm}}$)
and the collapses are given by  $\ket{\nu_\al} \to
\ket{\vartheta_1,{+}} \to \ket{\nu_\be} \to \ket{\vartheta_1,{-}}
\to \ket{\nu_\al} $. As long
 as the collapses lead   to closed loop encircling the origin, we will
obtain this topological phase.
So this phase of $\pi$ appears whether we consider vacuum and/or
ordinary matter with constant density or  with slowly changing (but
noncyclic) electron number density. This is due to the topological
character of this phase, which will be preserved as long as we have
$CP$-conserving (real) Hamiltonian and states are always lying on a
great circle in the $x-z$ plane in the \poi sphere.

  Next we write down an explicit
expression for the observable quantities, {\it i.e.} appearance
 and survival probabilities for two neutrino flavors.
Using the general expression obtained in Eq.~\ref{prob}, the
appearance probability for transition $\nu_e \to \nu_\mu$ is given
by~\footnote{In order to connect with the standard expressions used
in neutrino literature, we shall revert to $\Theta$ instead of
$\vartheta/2$.} \bea {\cal P} (\nu_e \to \nu_\mu) &=&
\mathbb{U}_{e+}^\star (\Theta_1) \mathbb{U}_{\mu+}^{} (\Theta_2)
\mathbb{U}_{\mu +}^{\star}
(\Theta_2) \mathbb{U}_{e+}^{} (\Theta_1) \nonumber\\
&+ & \mathbb{U}_{e-}^\star (\Theta_1) \mathbb{U}_{\mu-}^{}
(\Theta_2) \mathbb{U}_{\mu -}^{\star} (\Theta_2) \mathbb{U}_{e-}^{}
(\Theta_1) \nonumber\\
&&
\!\!\!\!\!\!\!\!\!\!\!\!\!\!\!\!\!\!\!\!\!\!\!\!\!\!\!\!\!\!\!\!\!\!\!\!\!\!\!\!\!
+~[\mathbb{U}_{e-}^\star (\Theta_1) e^{i  {\cal D}_- }
\mathbb{U}_{\mu -}^{} (\Theta_2) \mathbb{U}_{\mu +}^{\star}
(\Theta_2) e^{-i  {\cal D}_+ } \mathbb{U}_{e+}^{} (\Theta_1) +
{\rm{c.c.}} ].\nonumber\\
 \eea
Note that the matrix $\mathbb{U} (\Theta)$ is the lepton mixing
matrix (defined in a basis where the charged lepton mass matrix is
diagonal). It is also referred to as the
Pontecorvo-Maki-Nakagawa-Sakata (PMNS) matrix~\cite{pontecorvo,mns}
and connects the flavor states to the mass eigenstates. For the
$2\times 2$ case, it is a real orthogonal rotation matrix given by
\beq \mathbb U (\Theta) =
\begin{pmatrix}
\cos \Theta & \sin \Theta \\ -\sin \Theta & \cos \Theta
\end{pmatrix}~. \eeq
 Substituting the
elements of $\mathbb{U} (\Theta)$ we get \bea {\cal P} (\nu_e \to
\nu_\mu) &=& \cos^2 \Theta_1 \sin ^2 \Theta_2 + \sin^2 \Theta_1 \cos
^2 \Theta_2 \nonumber\\ &&
\!\!\!\!\!\!\!\!\!\!\!\!\!\!\!\!\!\!\!\!\!\!\!\!\!\!\!\!\!\!\!\!\!\!\!\!\!\!\!\!\!\!\!\!\!\!
+~ [2 \cos ({\cal D}_+ - {\cal D}_-) ] (-\sin \Theta_1) \cos\Theta_2
\sin \Theta_2 \cos \Theta_1~. \label{eqcos}  \eea  We note that
there are four inner products appearing in
 the interference term in the  final expression for the
 probability  out of which the first three inner
  products, viz.,
$\scx{\vartheta_1,{+}}{\nu_e} = \mathbb{U}_{e+}(\Theta_1) = \cos
\Theta_1 > 0$,
$\scx{\nu_\mu}{\vartheta_2,{+}}= \mathbb{U}^\star_{\mu+}(\Theta_2)=
\sin \Theta_{2} > 0$
and $\scx{\vartheta_2,{-}}{\nu_\mu} = \mathbb{U}_{\mu -}(\Theta_2) =
\cos \Theta_2 > 0$
clearly implying that these states are mutually parallel to each
other  in pairs according to Pancharatnam's rule, which is to have
the inner product of any two states real and positive, while the
last one, $\scx{\nu_e}{\vartheta_1,{-}} =
\mathbb{U}_{e-}^\star(\Theta_1) = - \sin \Theta_1 < 0$ by
Pancharatnam's rule  has $\ket{\nu_e}$  antiparallel to
$\ket{\vartheta_1,{-}}$,  since the physically allowed values for
the mixing angles $\Theta_1$ and $\Theta_2$ are within  the interval
$[0,\pi/2]$ for $\delta m^2>0$~\cite{yossi} (On the \poi sphere, the
corresponding $\vartheta_1$ and $\vartheta_{2}$ can take values
between $[0,\pi]$).  The minus sign appearing in the interference
term is thus the Pancharatnam's phase of $\pi$ appearing in the
neutrino oscillation formula (see Fig.~\ref{fig3}(a)).

If in a hypothetical situation, for some range of parameters
$\Theta_{1}$ and $\Theta_2$, the first three of the inner products
are real and negative ({\it i.e.} the states are aligned
antiparallel to each other or completely out of phase), while the
fourth inner product is real and positive (the states are in phase)
then also we will have this minus sign. The nontransitivity also
holds here leading to the non-trivial topological phase of $\pi$.
This situation where the inner product becomes real and negative
defines an ``antiparallel" rule (in the same spirit in which
Pancharatnam defined his rule of two states being ``in phase or
parallel") would correspond to the norm of the vector sum of the two
states being at its minimum value. Physically, this implies the
interference of the two given states would be destructive and the
resulting state will have minimum intensity or a dark fringe in
optics.

The existence of Pancharatnam's phase of $\pi$ can be simply
connected to the fact that the mixing matrix $\mathbb{U} (\Theta)$
matrix for two flavors is an orthogonal rotation matrix
parameterized by the mixing angle $\Theta$ of which one element has
a negative sign. Thus, this phase is {\it{built into}} the structure
of ${\mathbb{U}} (\Theta)$ matrix.

The survival probability is given by \bea {\cal P} (\nu_e \to \nu_e)
&=& \cos^2 \Theta_1 \cos ^2 \Theta_2 + \sin^2 \Theta_1 \sin ^2
\Theta_2 \nonumber\\ &&
\!\!\!\!\!\!\!\!\!\!\!\!\!\!\!\!\!\!\!\!\!\!\!\!\!\!\!\!\!\!\!\!\!\!\!\!\!\!\!\!\!
+~ [2 \cos ({\cal D}_+ - {\cal D}_-) ] \sin \Theta_1 \cos\Theta_2
\sin \Theta_2 \cos \Theta_1~. \label{pee} \eea Note that in the case
of survival probability, the cross term does not pick up any nonzero
topological phase, and geometrically this is exactly what we had
expected from Fig~\ref{fig3}(b). The loop in $\vartheta$-space is
open in this case, and this is what leads to this result. The
topological phase of the interference term in survival probability
is zero, while it is $\pi$ in the case of the appearance
probability, and this fact is in accord with unitarity.

The above expressions (Eqs.~\ref{eqcos} and \ref{pee}) reduce to the
standard results~\cite{kuo,walecka,yossi,kim} for vacuum if we
substitute $\Theta_1 = \Theta_2 = \Theta$, \bea {\cal P} (\nu_e \to
\nu_\mu) &=& \sin^2 2 \Theta \sin^2 \dfrac{\delta m^2 l}{4 E} \quad
{\rm{and}}
\nonumber\\
{\cal P} (\nu_e \to \nu_e) &=& 1- \sin^2 2 \Theta \sin^2
\dfrac{\delta m^2 l}{4 E}~, \label{vacprob}
 \eea
where in the ultrarelativistic limit, we can use  $t \simeq l$ and
$p \simeq E$ leading to  ${\cal D}_{\pm}=\pm \delta m^2 l/2E$ (see
Eq.~\ref{correctadiab}) for the vacuum case ($V_C=V_N=0$).
In constant density matter, the quantities $\Theta$ and ${\delta
m}^2$ in Eq.~\ref{vacprob} are replaced by their respective
renormalized values in matter, $\Theta^m$ and $({\delta m}^2)^m$ but
the form of the expression will remain the same. Hence our result is
consistent with the standard neutrino oscillation formulation, and
it provides a clear geometric interpretation of the phenomenon of
neutrino oscillations. More precisely, the standard result for
neutrino oscillations is in fact a realization of the Pancharatnam
topological phase.

 \vskip .6 true cm
\section{Discussion}
\label{disc}

As mentioned in the introduction, the existing work on the subject
of geometric phases in neutrino oscillations led to the widespread
belief that the two flavor neutrino oscillation formulae in CP
conserving situations were devoid of any geometric or topological
phase component. Appearance of the cyclic Berry phase was dismissed
on the grounds of not having any time-varying parameter in vacuum
and having only one {\it{essential}} parameter (thereby enclosing no
area) in the case of normal
matter~\cite{nakagawa,Naumov:1991,Naumov:1991rh,Naumov:1992,Naumov:1993vz,he}.
Concerning the appearance of the general geometric phase in the two
flavor neutrino case for propagation in vacuum, there are claims
reporting its appearance~\cite{Blasone:1999tq,wang}. But, it should
be noted that such terms appeared only at amplitude level and as
argued earlier, a phase appearing in the amplitude can be observed
only via a split-beam experiment, which is not feasible to design in
the case of neutrinos.

In this paper, we have examined the minimal case of two flavor
neutrino oscillations and $CP$ conservation. Contrary to all
existing claims in the literature concerning  the geometric or
topological phase in two flavor neutrino oscillation probabilities,
our study provides the first clear prediction that a topological
phase of $\pi$ exists at the probability level even in the minimal
case of $CP$ conservation. We show that it is inherently present in
the physics of neutrino oscillations via the structure of the PMNS
neutrino mixing matrix. This existence of this topological phase is
linked to the presence of a flux line  of strength $\pi$ at the
origin of ray space, which is connected to the degeneracy point
associated with the null Hamiltonian.

Pancharatnam's idea  is quite useful in terms of predictive power as
it allows for a clear visualization of the appearance of such a
phase due to geometric effects without doing any  algebra. Our
prescription is  general as it contains effects due to collapses and
also due to adiabatic evolution. In the absence of either of these,
one would get the same phase. So no matter what the details are, as
long as the singular (degeneracy) point is enclosed by a cyclic loop
(in the space of rays) as $\vartheta$ is varied from $0 \to 2\pi$,
we will get this phase, and this is due to its topological
robustness. The adiabatic and collapse processes both conspire in
such a fashion that the net phase would always be $\pi$. This does
not happen for geometric phases.

 The topological phase obtained in this paper is a consequence of
 anholonomy,
 which can arise in situations even when there is no curvature. The
 most
 striking example of this is the Aharonov-Bohm effect~\cite{shaperebook}.
 To experience the effect of anholonomy, the main requirement is to
 encircle the singular point, this fact was exploited by Herzberg
 and Longuet-Higgins in pointing out the topological phase in molecular physics.
 On the other hand, for Berry's phase to appear, a net curvature is a must which is fulfilled
 by having at least two essential parameters  in
the Hamiltonian varying cyclically.
 This is an important distinction between the geometric phases as obtained by Herzberg and
 Longuet-Higgins and by Berry.

If we consider mixed flavor states~\footnote{Here mixed state refers
to a superposition of pure flavor states and should not be confused
with the mixed states in the density matrix language which are not
pure.} instead of the pure flavor states, there will be a greater
number of physical situations (or, possible diagrams for the
interference term like the ones shown in Fig.~\ref{fig3} for pure
flavor states) that can be explored to see if one encircles the
singular point or not. A mixed state corresponds to a general point
on the surface of the \poi sphere like an elliptically polarized
state in optics. If the mixed states are such that they lie on the
$x-z$ plane, it will always lead to the same quantized topological
phase of $\pi$. But, for a general mixed state lying anywhere else
on the \poi sphere, the phase will be geometric in nature.

It might be a nontrivial task to extend our geometrical
interpretation to the case of three neutrinos flavors because it
will involve a higher dimensional sphere (the ray space is
${\mathcal {CP}}^{2}$ for the three level quantum system).

 It is natural to ask what happens when we invoke $CP$ violation. In
vacuum, $CP$ violation cannot be induced in the two flavor case as a
consequence of  $CPT$ invariance and
unitarity~\cite{Akhmedov:2004ve}. However, matter with constant or
varying density can induce $CP$ violation via the coherent forward
scattering of neutrinos with background matter. If we introduce $CP$
violation induced by background matter with constant
density~\cite{Akhmedov:2004ve}, we still expect to get the same
phase of $\pi$ as we have two pairs of orthogonal states that will
always lie on a great circle. If the density is varying slowly
(adiabatic condition holds), then the intermediate states (connected
by adiabatic evolution) will be lifted from the great circle, hence
resulting in a path-dependent solid angle, and the phase will be
geometric~\cite{pmcpv}.

\acknowledgements{The author is deeply indebted to Joseph Samuel and
Supurna Sinha for numerous useful discussions leading to the present
work and critical comments on the manuscript. Support from the
Weizmann Institute of Science, Israel during the initial stages of
this project  is gratefully acknowledged. }

\bibliographystyle{apsrev}
\bibliography{referencesberry}

\end{document}